\documentclass[pra,twocolumn,showpacs]{revtex4}
\usepackage{amsmath}
\usepackage[dvips,xdvi]{graphicx}
\usepackage{amssymb}
\usepackage{epsfig}
\usepackage{color}
\graphicspath{{../figures/}}

\begin{document}
\title{Velocity selector with a microwave magnetic dipole transition}

\author{L. O. Casta\~nos$^{*1,2}$ and E. Gomez$^{+2}$}

\affiliation{1. Departamento de F\'{i}sica Matem\'{a}tica, Instituto de Investigaciones en Matem\'{a}ticas Aplicadas y en Sistemas, Universidad Nacional Aut\'{o}noma de M\'{e}xico, C.P. 04510, M\'{e}xico\\
             2. Instituto de F\'{i}sica, Universidad Aut\'{o}noma de San Luis Potos\'{i}, San Luis Potos\'{i} 78290}
\begin{abstract}
We proposed a method based on microwave magnetic dipole transitions to prepare samples of atoms with well defined position and velocity. Each microwave pulse corresponds to a position measurement for the atoms and two pulses separated by a given delay result in a velocity measurement. The method gives velocity sensitivity approaching that obtained with Raman transitions but it is easier to implement. Moreover, it has the advantages that it also selects in position and has less demanding experimental requirements. The method can be demonstrated in a magneto-optical trap.
\end{abstract}

\maketitle

An important tool in atomic physics is the capability of velocity measurements. The simplest way to measure velocity is by time of flight, something that is heavily used, for example, to determine the momentum distribution in a Bose-Einstein condensate \cite{anderson95,davis95,lett88} or to reconstruct a molecular wavefunction by looking at the expanding products after dissociation \cite{schmidt12,leventhal67}. More precise measurements of velocity exploit the Doppler effect, since it introduces a frequency shift proportional to the velocity \cite{kasevich91}. The careful measurement of velocity (or velocity changes) offers the possibility to measure accelerations \cite{kasevich92,ferrari06}, rotations \cite{gustavson97}, fundamental constants \cite{battesti04,stuhler03}, and it is used to prepare atomic samples with a given velocity distribution \cite{berthoud99}.

Raman transitions between hyperfine levels are generally used for fine velocity selection \cite{kasevich91}. Microwave magnetic dipole (M1) transitions are not used for that purpose because the momentum of a microwave photon is orders of magnitude smaller than that of an optical photon. Still M1 transitions are simpler to implement and have longer coherence times compared to Raman transitions \cite{dudin13}. We present a way to use M1 transitions for fine velocity selection that reaches competitive sensitivity with respect to Raman transitions and has position sensitivity as well \cite{castanos13}, something useful to prepare atomic samples with well defined phase space distributions \cite{castanos13}.

Consider atoms moving vertically in a static magnetic field $B_{\mbox{\tiny ST}}(z) = \eta z$ along the vertical $z$-axis. By turning on a microwave magnetic field we drive transitions between magnetic sensitive states only for atoms located at a particular position. Two position measurements separated by a time interval $\Delta t$ give the selection in velocity along the $z$-axis. Each position measurement corresponds to a microwave $\pi$-pulse followed by a cleaning laser pulse that removes the atoms not excited by the transition. The position measurement is analogous to that used in Nuclear Magnetic Resonance (NMR) \cite{NMR}, the difference being that now the atoms continue moving due to the low densities used in laser cooling and a second position measurement provides the velocity selection.

The Hamiltonian of the system (alkali atom + magnetic field) along the $z$-axis and in the dipole and long-wavelength approximations is \cite{castanos13}
\begin{equation}
H(t) \ = \ \frac{1}{2M}P_{z}^2  + Mg_{0}Z + H_A -\mbox{\boldmath$\mu$}\cdot\left[ \eta Z \mathbf{z} + \mathbf{B}_{p}(t) \right] \ ,
\label{Hsystem}
\end{equation}
where $M$ is the mass of the atom, $g_{0} = 9.8$ m/s$^{2}$ is the gravitational acceleration, $H_{A}$ is the hyperfine Hamiltonian, $Z$ and $P_{z}$ are the center of mass position and momentum operators, respectively, $\mbox{\boldmath$\mu$}$ is the magnetic dipole moment operator of the atom, and $\mathbf{B}_{p}(t)$ is the microwave magnetic field. We only need a 1D Hamiltonian since the Schr\"{o}dinger equation is separable for the magnetic field considered \cite{castanos13}. We consider transitions between hyperfine \textit{stretched states} $\left| F = F_{\scriptscriptstyle{-}}, M_{\scriptscriptstyle{F}} = \pm F_{\scriptscriptstyle{-}} \right\rangle \leftrightarrow \left| F=F_{\scriptscriptstyle{+}}, M_{\scriptscriptstyle{F}} = \pm F_{\scriptscriptstyle{+}} \right\rangle$ with $F_{\pm}= I \pm 1/2$, since they have the strongest sensitivity to the magnetic field. We assume a nuclear spin $I\geq 1/2$. To drive the transitions, the microwaves $\mathbf{B}_{p}(t)$ propagate along the positive $x$-axis with linear polarization along the positive $y$-axis. 

For the internal degrees of freedom we use a position dependent basis composed of the eigenstates of $H_A -\mbox{\boldmath$\mu$}\cdot \eta z \mathbf{z}$ \cite{castanos13}: 
\begin{eqnarray}
\label{BreitRabi}
\left( H_{A} -\eta \mu_{z}z\right) \left| F,M_{\scriptscriptstyle{F}} \right\rangle \ = \ V_{\scriptscriptstyle{F,M_{\scriptscriptstyle{F}}}}(\kappa z)\left| F,M_{\scriptscriptstyle{F}} \right\rangle \ .
\end{eqnarray}
The eigenvalues are given by the Breit-Rabi formula \cite{castanos13,Tinkham}:
\begin{eqnarray}
\frac{V_{\scriptscriptstyle{F_{\scriptscriptstyle{-}},\pm F_{\scriptscriptstyle{-}}}}(\kappa z)}{\hbar \Delta W} &=& \mp F_{\scriptscriptstyle{-}}\gamma_{2}\kappa z - \frac{1}{2}\sqrt{1 \pm 2\frac{F_{\scriptscriptstyle{-}}}{F_{\scriptscriptstyle{+}}}\kappa z + (\kappa z)^{2}} \ , \cr
\frac{V_{\scriptscriptstyle{F_{\scriptscriptstyle{+}}}, \pm F_{\scriptscriptstyle{+}}}(\kappa z)}{\hbar \Delta W} &=& \frac{1}{2} \pm \gamma_{1}\kappa z \ ,
\end{eqnarray}
with $\hbar \Delta W$ the hyperfine splitting, $\gamma_{2} = g_{\scriptscriptstyle{I}}\mu_{\scriptscriptstyle{N}}/\mathsf{g}$, $\gamma_{1} = (g_{s}\mu_{\scriptscriptstyle{B}}-2Ig_{\scriptscriptstyle{I}}\mu_{\scriptscriptstyle{N}})/(2\mathsf{g})$,  $\mathsf{g} = (g_{s}\mu_{\scriptscriptstyle{B}}+g_{\scriptscriptstyle{I}}\mu_{\scriptscriptstyle{N}})$, $g_{s}$ and $g_{\scriptscriptstyle{I}}$ the electron and nuclear $g$-factors, and $\mu_{\scriptscriptstyle{B}}$ and $\mu_{\scriptscriptstyle{N}}$ the Bohr and nuclear magnetons, respectively. Also, $1/\kappa$ is the characteristic length of the system with
\begin{equation}
\kappa = \frac{\mathsf{g}}{\hbar \Delta W}\eta \ . \label{eqkappa}
\end{equation}
The use of the position dependent basis introduces couplings between ground state levels that are proportional to the following (non-dimensional) perturbation parameter \cite{castanos13}
\begin{equation}
\label{epsiloneq}
\epsilon = \frac{\hbar^{2}}{2M}\frac{\kappa^2}{\hbar \Delta W}.
\end{equation}
For all alkali atoms $\epsilon$ is very small, for example, $\epsilon = 8.9 \times 10^{-21}$ for $^{87}$Rb with $\eta = 25$ Gauss/cm.

The Sch\"{o}dinger equation is reduced to that of a two-level system in the rotating wave approximation \cite{castanos13}. To leading order in $\epsilon$ the atoms that get transferred between hyperfine levels with a probability $\geq 1/2$ (hereby known as the \textit{selected atoms}) lie in a position band of width approximately given by \cite{castanos13}
\begin{eqnarray}
\label{poswidth}
\Delta_{z} \ \simeq \ \left. \frac{2\hbar \Omega_{\scriptscriptstyle{R}}(z) }{\left\vert \frac{d}{dz}\left[ V_{\scriptscriptstyle{F_{\scriptscriptstyle{+}}}, \pm F_{\scriptscriptstyle{+}}}(\kappa z) - V_{\scriptscriptstyle{F_{\scriptscriptstyle{-}}}, \pm F_{\scriptscriptstyle{-}}}(\kappa z) \right] \right\vert} \right\vert_{z =\langle Z \rangle (t_{0})},
\end{eqnarray}
with $\Omega_{\scriptscriptstyle{R}}(z)$ the (angular) Rabi frequency of the $\pi$-pulse at the position $z$, $t_{0}$ the instant in which the $\pi$-pulse is applied, and $\langle Z \rangle (t_{0})$ the expected value of the position of the atom at time $t_{0}$. The dependence on position of $\Omega_{\scriptscriptstyle{R}}(z)$ arises mainly from a position dependent detuning with a small atom-$\mathbf{B}_{p}(t)$ coupling variation \cite{castanos13}. Explicitly,
\begin{eqnarray}
\Omega_{\scriptscriptstyle{R}}(z) &=& \sqrt{ \delta (z)^{2} + 4\vert \Omega_{0}(z) \vert^{2}} \ ,
\end{eqnarray}
with the detuning $\delta (z)$ and coupling $\Omega_{0}(z)$ given by
\begin{eqnarray}
\hbar\delta (z) &=& V_{\scriptscriptstyle{F_{+},\pm F_{+}}}(\kappa z) - V_{\scriptscriptstyle{F_{-},\pm F_{-}}}(\kappa z) - \hbar \omega_{A} \ ,\cr
\hbar\Omega_{0}(z) &=& -\frac{B_{0}}{2}\langle F_{+},\pm F_{+} \vert \mu_{y} \vert F_{-},\pm F_{-}  \rangle \ .
\end{eqnarray}
Note that $\mathbf{B}_{p}(t) = \mathbf{y}B_{0}\mbox{cos}[\omega_{A}(t-t_{0})]$ and that $\vert F_{-}, \pm F_{-} \rangle$ depends on $z$.

Equation (\ref{poswidth}) is accurate as long as the atom is well localized around its expected value of position and it does not move much during the $\pi$ pulse \cite{castanos13}, that is, if
\begin{eqnarray}
1 &\gg& \epsilon \left\vert \left[ \frac{\langle P_{z} \rangle (t_{0})}{\hbar\kappa} \right]^{2} + \frac{1}{2[ \kappa^{2}\Delta Z(0)^{2} + i\epsilon \Delta W t_{0} ]}  \right\vert \times \cr 
&& \qquad \times\frac{\pi \Delta W}{\Omega_{\scriptscriptstyle{R}}[\langle Z \rangle (t_{0})]}\left[ 2\pi \kappa^{2}\Delta Z (t_{0})^{2} \right]^{-1/4} \ .
\end{eqnarray}
Here $\langle P_{z} \rangle (t_{0})$ is the expected value of the momentum of the atom at time $t_{0}$, $\Delta Z(0)$ is the initial root-mean-square (RMS) deviation of the position of the atom, and $\Delta Z(t_{0})$ is the RMS deviation of the position at time $t_{0}$. 

At low magnetic fields (that is, for $|\kappa z| \ll 1$) equation (\ref{poswidth}) reduces to
\begin{eqnarray}
\label{poswidth2}
\Delta_{z} \ \simeq \ \frac{\hbar \Omega_{\scriptscriptstyle{R}}[\langle Z \rangle(t_{0})] }{\mu_{\scriptscriptstyle{B}} \eta} \left( \frac{I+1/2}{I} \right).
\end{eqnarray}
The scaling in Eq.~\ref{poswidth2} can be easily recovered using $\delta \approx \Omega_{\scriptscriptstyle{R}}$ that corresponds to the approximate spectral width of the transition, with $\delta$ given by the Zeeman effect. The position sensitivity grows with $\eta$, since the resonant frequency changes faster with position. For a $10$ $\mu$s $\pi$-pulse and $^{87}$Rb atoms in a gradient of $\eta = 25$ Gauss/cm one gets from (\ref{poswidth2}) that $\Delta_{z} = 19$ $\mu$m. 

For a velocity measurement we apply a laser pulse to project the atoms to the corresponding level followed by a cleaning pulse to keep only the atoms that made the transition and we let them continue moving vertically for a time $\Delta t$. Then we apply another $\pi$-pulse at a different position adjusting the microwave frequency accordingly. The width of the velocity selection is given approximately by \cite{castanos13}
\begin{equation}
\label{velwidth}
\Delta_{v} \ \simeq \ \left( 2 \Delta_{z} \right) \frac{1}{\Delta t} \ .
\end{equation}
Using the value of $\Delta_{z} = 19$ $\mu$m above with $\Delta t = 28$ms one gets from (\ref{velwidth}) that $\Delta_{v} = 1.4$ mm/s, which starts approaching the results obtained using velocity selective Raman transitions \cite{chabe07}.

Velocity selection is usually done by driving Raman transitions \cite{kasevich91,kasevich92,gustavson97,battesti04,stuhler03}. In that case the selectivity is given by \cite{moler92}
\begin{equation}
\label{velwidthR}
\Delta v_{\scriptscriptstyle{R}} \simeq \left( \frac{1}{2k} \right) \frac{1}{\Delta t_{\scriptscriptstyle{R}}} \ , 
\end{equation}
where $\Delta t_{\scriptscriptstyle{R}}$ is the duration of the Raman pulse and $k$ is the wave number associated with the Raman transition. Microwave transitions are never considered for velocity measurements because of exactly this dependence on the photon momentum ($\hbar k$) which is several orders of magnitude smaller than with optical transitions. Equations~\ref{velwidth} and \ref{velwidthR} have a common dependence on the measurement duration with a different prefactor that in the Raman case depends on the wave number ($k$) and in the microwaves on the magnetic field gradient ($\eta$) and the Rabi frequency. It is important to notice that the requirements for frequency stability in the microwave method are relaxed compared to the Raman technique, since one uses two short pulses separated by a long time delay in the former as opposed to a single long pulse in the latter. The higher sensitivity of the Raman transitions is compensated by the simplicity and potentially longer coherence times of the microwave method.

To better understand the position and velocity selection consider a cloud of non-interacting alkali atoms in the state $\vert F_{-}, F_{-} \rangle$. The application of the first $\pi$ pulse at time $t=0$ selects the atoms near a particular position $z_{\scriptscriptstyle{0R}}$. The atoms continue moving vertically and the expected value of the position of each atom at later times follows the classical equation of motion and is given by \cite{castanos13}
\begin{eqnarray}
\label{freefall}
\langle Z \rangle (t) \ = \ z_{0} + \frac{p_{0}}{M} t - \frac{1}{2}g_{\mbox{\tiny eff}} t^{2} \ ,
\end{eqnarray} 
with $z_{0} \simeq z_{\scriptscriptstyle{0R}}$ and $p_{0} = \langle P \rangle (0)$ the initial expected values of position and momentum of the atom, respectively. Also, $g_{\mbox{\tiny eff}} = g_{0} + \eta \mathsf{g} \gamma_{1} /M$ is the effective acceleration felt by the atom (it includes both the gravitational acceleration and the effect of the internal energy of the atom placed in the static magnetic field). Finally, the second $\pi$-pulse at time $t=\Delta t$ selects atoms near a different position $\langle Z \rangle (\Delta t) \simeq z_{f}$.  

\begin{figure}
\center
\includegraphics[width=7cm]{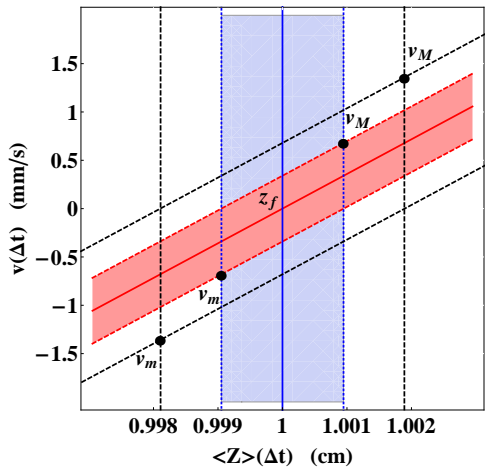}
\caption{Region of phase space selected by two $\pi$-pulses. For the figure we use a $^{87}$Rb atom in the internal state $\vert F_{-},F_{-}\rangle$ before the first $\pi$-pulse. Also, $\eta = 25$ Gauss/cm, $z_{\scriptscriptstyle{0R}}=0$, $z_{f}=1$ cm and $\Delta t = 28$ ms. The shaded areas corespond to the regions selected by 10 $\mu$s pulses whereas the resulting area for 5 $\mu$s pulses is delimited by the black-dashed lines.}
\label{bandselection}
\end{figure}

Figure \ref{bandselection} shows the phase space diagram at the end of the second pulse with the bands introduced by each transition. The first pulse after the evolution given by (\ref{freefall}) produces the red shaded band delimited by the red-dashed lines in Fig.~\ref{bandselection}. The slope is $1/\Delta t$ and becomes more horizontal as $\Delta t$ increases. The second pulse produces the vertical blue shaded band delimited by the blue-dashed lines, since we are describing the phase space distribution immediately after the position measurement. The crossing of the two shaded bands gives the selected region of phase space centered at $z_{f}$ and $v_f=(z_f-z_{\scriptscriptstyle{0R}} - g_{\mbox{\tiny eff}} \Delta t^{2}/2)/\Delta t$ (which is zero in our case since we consider the maximum height of the trajectory). Projecting the distribution onto the $v$-axis gives the marginal distribution of the velocity that has a width approximately equal to the distance between $v_{\scriptscriptstyle{M}}$ and $v_{\scriptscriptstyle{m}}$ and given in (\ref{velwidth}). 

The acceleration of the atoms depends on the hyperfine level due to their different response to the static magnetic field \cite{castanos13}. This separates the atoms that made the transition from the rest. A similar thing happens with velocity dependent Raman transitions, except that, instead of a different acceleration, it is a velocity difference what separates the atoms. 

Measurements of the complete phase space distribution are possible with our method and not only the marginal velocity distribution as with traditional velocity dependent Raman transitions. This is the case because the two measurements select a particular region of phase space (Fig.~\ref{bandselection}). On the other hand, for velocity only measurements our method has a reduced signal compared to the Raman technique. For example, for atoms in a magneto optical trap with a typical size of 1 mm, a position selection of $\Delta_{z} = 19$ $\mu$m as before would reduce the sample size for velocity only measurements by two orders of magnitude. Raman transitions with magnetic sensitivity have also been demonstrated \cite{terraciano07}.

The results (\ref{poswidth}) and (\ref{velwidth}) where obtained assuming non-interacting atoms so that the evolution between pulses given in (\ref{freefall}) is valid. This is usually the case even for Bose-Einstein condensates after a few ms of free expansion. It is also required that the atoms are well-localized around their expected value of position. To understand this requirement suppose that the initial state of an atom is a pure and separable state of the center of mass and internal degrees of freedom $\vert \psi (0) \rangle = \vert \psi_{\mbox{\tiny cm}}(0) \rangle \otimes \vert F_{+}, \pm F_{+} \rangle$. The state of the atom at later times is still separable and of the form $\vert \psi (t) \rangle = \vert \psi_{\mbox{\tiny cm}}(t) \rangle \otimes \vert F_{+}, \pm F_{+} \rangle$ \cite{castanos13}. In particular, if $\vert \psi_{\mbox{\tiny cm}}(0) \rangle$ is a coherent state, then the wavefunction $\psi_{\mbox{\tiny cm}}(z, t) = \langle z \vert \psi_{\mbox{\tiny cm}}(t)\rangle$ is a Gaussian state whose RMS deviation $\Delta Z(t)$ is exactly the same as that of a freely evolving Gaussian wavepacket, whereas the deviation in momentum $\Delta P(t)$ remains constant since the force is independent of $z$ \cite{castanos13}. The transition probability is \cite{castanos13}
\begin{eqnarray}
\label{Prob}
P(t) &\simeq& \int_{-\infty}^{+\infty}dz \ \vert \psi_{\mbox{\tiny cm}} (z,t_{0}) \vert^{2} \times \cr
&& \times \left[ \frac{2\vert \Omega_{0}(z) \vert}{\Omega_{\scriptscriptstyle{R}}(z)} \right]^{2} \mbox{sin}^{2}\left[ \frac{\Omega_{\scriptscriptstyle{R}}(z)}{2}(\tau-\tau_{0}) \right] \ . \ \ \
\end{eqnarray}
This corresponds to the well known formula describing Rabi oscillations but averaged over the position probability density function of the atom $\vert \psi_{\mbox{\tiny cm}}(z,t_{0})\vert^{2}$ at the time of the application of the microwave pulse. An extended wave packet leads to the reduction of both the probability of transition in a $\pi$-pulse and the visibility of Rabi oscillations \cite{castanos13}. The result is that the selected atoms do not make the transition efficiently.

The selected position band $\Delta_{z}$ can be increased by reducing the pulse duration (or increasing $\Omega_{\scriptscriptstyle{R}}$) at the price of reducing the sensitivity  \cite{castanos13}. Still velocities around 1 mm/s are possible which are already a fraction of the recoil velocity.

We illustrate these facts with a $^{87}$Rb atom initially in the internal state $\vert F_{-}, F_{-}\rangle$. We use the same conditions as before with $\eta = 25$ Gauss/cm and $\Delta t = 28$ ms to reach a maximum height of $1$cm. We take an RMS deviation in the position of the atom of $\Delta Z(0) = 3$ $\mu$m after the first $\pi$ pulse and at maximum height it becomes $\Delta Z (\Delta t) = 4.5$ $\mu$m. The two 10 $\mu$s $\pi$ pulses select the shaded area in Fig.~\ref{bandselection} as explained before. The wave packet localization here is not good enough since the initial wave packet size gives a transition probability (Eq.~\ref{Prob}) of 0.91 for the first pulse and 0.82 for the second one. The second pulse is less efficient due to the expansion of the wave packet. In order to have a more efficient transfer the duration of the pulses needs to be decreased. Taking 5 $\mu$s pulses gives a transition probability of 0.98 and 0.95 for the first and second pulses respectively. The smaller pulse also makes the phase space selection less sensitive as it is shown by the black-dashed lines in Fig.~\ref{bandselection}. Here the position selection is 38 $\mu$m and the velocity selection is 2.7 mm/s.

We now discuss some aspects for the experimental demonstration of velocity selection using microwave transitions. The static magnetic field can be implemented with two parallel circular coils of radius $R$ carrying steady currents $I_{\scriptscriptstyle{c}}$ in opposite directions with centers in $z = \pm d_{0}$ \cite{castanos13}. The magnetic field produced by the two coils near their axis and far away from them and sufficiently far away from the origin is aproximately given by $\mathbf{B}_{\mbox{\tiny 2C}}(\mathbf{r}) = \eta z \mathbf{z}$ \cite{castanos13}. The field is readily available in a magneto-optical trap as long as the displacement along $z$ from the origin is larger than the perpendicular one from the axis. Rather than moving the atoms it is possible to instead shift the zero of the field gradient by adding a bias field. Although magneto-optical traps normally use small gradients ($\eta \sim 5$ Gauss/cm  \cite{valenzuela12}), higher values of $\eta$ can be achieved by increasing the current or going to a magnetic trap.

To obtain efficient excitation the detuning needs to be stable to $\delta<\Omega_{\scriptscriptstyle{R}}$. For the 10 $\mu$s $\pi$ pulse considered above with $\eta=25$ Gauss/cm, that means that the bias field needs to be stable to about 0.2 Gauss and the magnetic field gradient to one part in $10^{3}$ for a 1 cm displacement. Both of these requirements are easy to achieve. The stability on the microwave frequency is also not demanding and since each $\pi$ pulse is relatively short there is no need to sweep the frequency to stay on resonance. The frequency only needs to be adjusted between pulses.

In summary we have demonstrated that it is possible to do selection in both position and velocity with microwave magnetic dipole (M1) transitions. To show this we determined the evolution of an alkali atom in a magnetic field gradient and in the presence of the microwaves \cite{castanos13}. We derived simple formulas that characterized the selections in position and velocity and concluded that the proposed method has a sensitivity approaching that of velocity dependent Raman transitions \cite{castanos13}. Moreover, it has the advantages that it is simpler to implement and that it selects both in position and in velocity. The method should be directly applicable in a wide variety of situations where one has atoms already moving vertically. In particular, the proposed scheme can be readily demonstrated in a magneto-optical trap.

{\bf Acknowledgments}

We thank CONACYT and the UNAM for support. We thank V\'{i}ctor Valenzuela from UAS for fruitful discussions.
\\
\\
* LOCCJ@yahoo.com\\
+ egomez@ifisica.uaslp.mx

\end{document}